\documentclass[10pt, conference, compsocconf]{IEEEtran}
\usepackage{cite}
\ifCLASSINFOpdf
   \usepackage[pdftex]{graphicx}
\else
\fi
\usepackage[cmex10]{amsmath}
\usepackage{ amssymb }
\usepackage{algorithmic}
\usepackage{float}
\usepackage{hyperref}
\newcommand\AIive{\textit{AIive}}
\begin{document}
% can use linebreaks \\ within to get better formatting as desired
\title{AIive: Interactive Visualization and Sonification of Neural Networks \\ in Virtual Reality}

\author{
\IEEEauthorblockN{Zhuoyue Lyu}
\IEEEauthorblockA{Department of Computer Science\\
University of Toronto\\
Toronto, Canada\\
zhuoyue@dgp.toronto.edu
}
\and
\IEEEauthorblockN{Jiannan Li}
\IEEEauthorblockA{Department of Computer Science\\
University of Toronto\\
Toronto, Canada\\
jiannanli@dgp.toronto.edu
}
\and
\IEEEauthorblockN{Bryan Wang}
\IEEEauthorblockA{Department of Computer Science\\
University of Toronto\\
Toronto, Canada\\
bryanw@dgp.toronto.edu
}
}

% \author{\IEEEauthorblockN{Zhuoyue Lyu, Jiannan Li, Bryan Wang}
% \IEEEauthorblockA{Department of Computer Science\\
% University of Toronto\\ \{zhuoyue, jiannanli, bryanw\}@dgp.toronto.edu}}

\maketitle

\begin{figure}[!t]
\centering
\includegraphics[width=\linewidth]{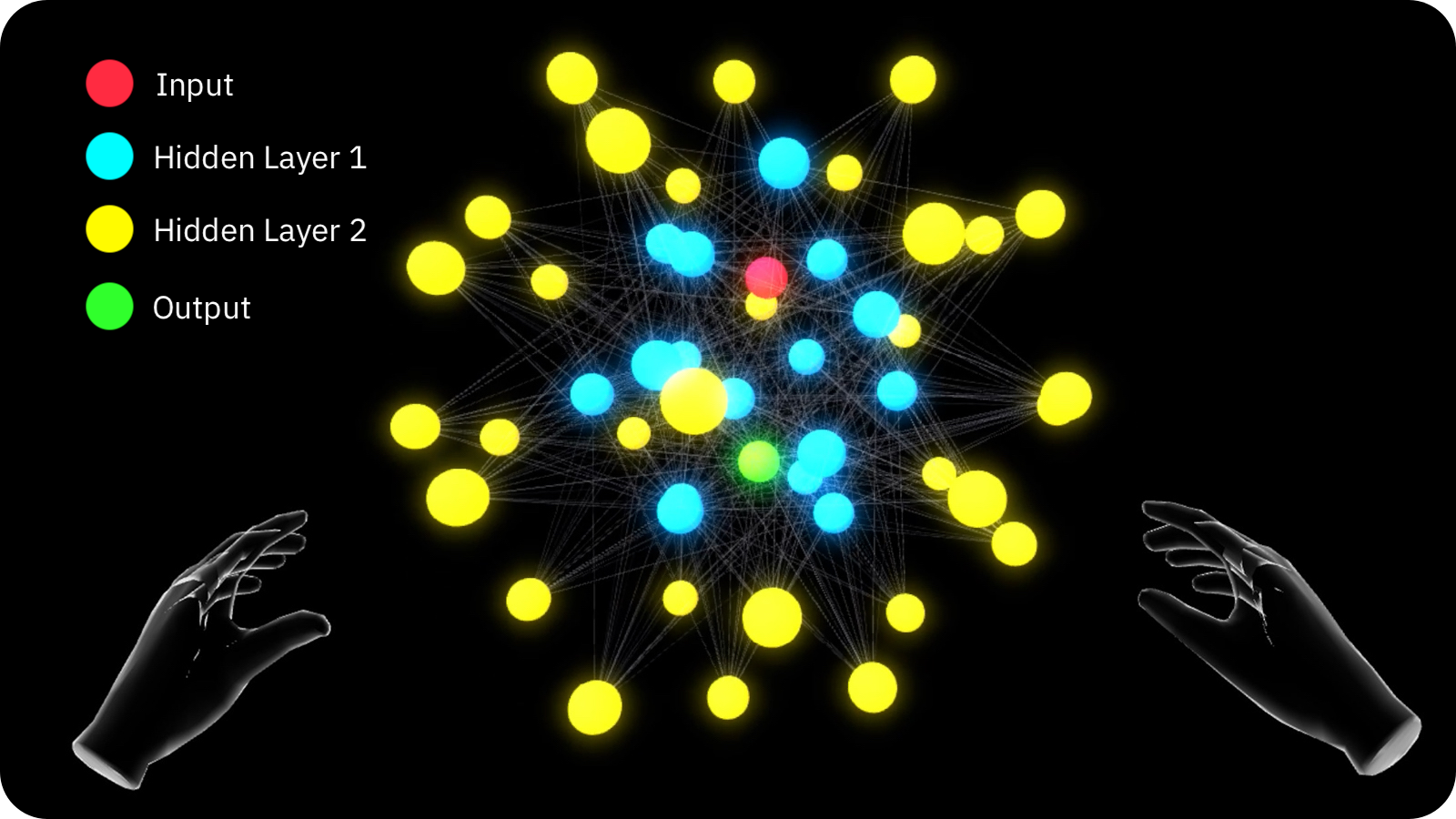}
\caption{The \AIive{} system, with the neural network at the center visualized as an interactive force-directed graph. \href{https://www.zhuoyuelyu.com/aiive}{\textbf{zhuoyuelyu.com/AIive}}}
\label{fig_sim}
\vspace{-0.2cm}
\end{figure}

\begin{abstract}
Artificial Intelligence (AI), especially Neural Networks (NNs), has become increasingly popular. However, people usually treat AI as a tool, focusing on improving outcome, accuracy, and performance while paying less attention to the representation of AI itself. We present \AIive{}, an interactive visualization of AI in Virtual Reality (VR) that brings AI ``alive". \AIive{} enables users to manipulate the parameters of NNs with virtual hands and provides auditory feedback for the real-time values of loss, accuracy, and hyperparameters. Thus, \AIive{} contributes an artistic and intuitive way to represent AI by integrating visualization, sonification, and direct manipulation in VR, potentially targeting a wide range of audiences.
\end{abstract}


\section{Introduction}
% Zhuoyue: the reviewers said the motivation if missing, so I added
AI has led to breakthroughs in areas such as image classification, object detection, and machine translation~\cite{zhang2021ai}. However, most research treats AI as a tool to accomplish tasks, focusing on boosting outcome and performance, with little attention paid to exploring the representation of NNs itself~\cite{10.1007/978-3-319-27857-5_77,9319101,pmlr-v123-herrmann20a,Zell1994}. Meanwhile, the lack of transparency and interpretability of AI is a problem that needs to be addressed~\cite{8371286,8400040,10.1145/3236009}. In this paper, we explore the concept of representing AI as a living being to uncover the inner workings of NNs in an experienceable way: it moves, changes colors, creates sounds as training progresses, and even responds to the users’ interaction with it.

To instantiate this concept, we present \AIive{}, an interactive visualization that brings AI ``alive", it leverages basic human sensory and motor activities: seeing, listening, and grabbing and moving objects for an intuitive, artistic, and enjoyable experience. Given the benefits of immersive analytics~\cite{4287241,SYLAIOU2010243}, data visualization and sonification in 3D~\cite{fazi2016feed,kramer2010sonification,Papachristodoulou2014SonificationOL}, we extend the 2D implementation of \textit{Immersions}~\cite{pmlr-v123-herrmann20a}, and visualize NNs as 3D force-directed graphs in VR that allow users to experience the training of the model and manipulate its architecture using virtual hands. We also provide the sonification of the accuracy, loss, learning rate, and momentum for real-time hyperparameter tuning.

\section{Related Work}

\subsection{Explainable AI and Democratizing AI}
% Zhuoyue: reviewers said we should talk about explainable AI. and the "Democratising AI" came from my discussion with Prof. Fernanda Viégas last week.
The hope of improving AI systems' transparency and accessibility triggered the research of explainable AI~\cite{8400040,10.1145/3236009} and democratization of AI~\cite{+2020}. For example, the AutoAI/AutoML by IBM~\cite{10.1145/3379336.3381474} and H2O.ai~\cite{candel2016deep} automate the end-to-end AI lifecycle to save data scientists from the low-level coding tasks; The TensorFlow Playground~\cite{smilkov2017directmanipulation} and GAN Lab~\cite{8440049} by Google allows direct-manipulation on the in-browser visualizations to help non-experts learn NNs and GANs. Those tools either decrease the experts' workload or help non-experts learn AI, thus requiring numeracy and graph literacy. \AIive{} however, focus on making AI training experienceable, thus only relying on basic sensory activities: sight, hearing, and touch in VR, which would potentially reach a broader audience.

\subsection{Immersive Analytics and Exploration}
% Zhuoyue: because the reviewrs asked: why VR? why Immersion? here is the answer to that:
Immersion provides benefits such as increased spatial understanding, decreased information clutter~\cite{4287241}, and the ``sense of being there", which is closely associated with satisfaction and appealing experience~\cite{SYLAIOU2010243}. VR headsets have been used to provide immersion: \textit{DataHop} enables users to layout data analysis steps in VR~\cite{10.1145/3379337.3415878}; \textit{AeroVR} provides an immersive environment to aid the aerodynamic design~\cite{tadeja_seshadri_kristensson_2020}, and VanHorn et al. developed a deep learning development environment in VR~\cite{vanhorn2019deep}. VR has also been used to teach programming~\cite{6871829,9419195,9320876}, and students found it more user-friendly, engaging, and better for visualization concepts compared to the traditional web-based system~\cite{9419195}. Thus, we built \AIive{} in VR to leverage those benefits.

\subsection{Data Visualization and Sonification in 3D}
3D data visualization provides the expanded domain of sensibility~\cite{fazi2016feed}, enabling a more comprehensive understanding of presented information: ``data objects",  ``data sculptures" were built to make data experienceable; Game-like infographics from datasets were made for playable and engaging experiences~\cite{10.1145/1978942.1979193}. Previous studies have shown that people found 3D visualization to be more satisfied and have lower workloads than the 2D counterpart~\cite{10.1145/3170427.3188537}. Data sonification has been used in various fields such as social sciences~\cite{de1999sonification,10.1145/3429290.3429307}, arts~\cite{10.1145/1873951.1874219}, and health~\cite{r2019hearing}, it can enhance visual representations without creating information overload~\cite{kramer2010sonification}, thus suitable for conveying dynamical information~\cite{Papachristodoulou2014SonificationOL}. There have been attempts to combine sonification with 3D visualization for representing the connectome of the human brain~\cite{Papachristodoulou2014SonificationOL}, communicating sensor data in workspaces~\cite{6126903}, and assisting music composition~\cite{774840}. \AIive{} builds upon the node-link visualization~\cite{10.1007/978-3-319-27857-5_77,9319101,pmlr-v123-herrmann20a,Zell1994} and focuses on sonifying parameters and performances of NNs.

\section{AIive Interface}
\subsection{Visualization}
\begin{figure}[!htb]
\centering
\includegraphics[width=0.8\linewidth]{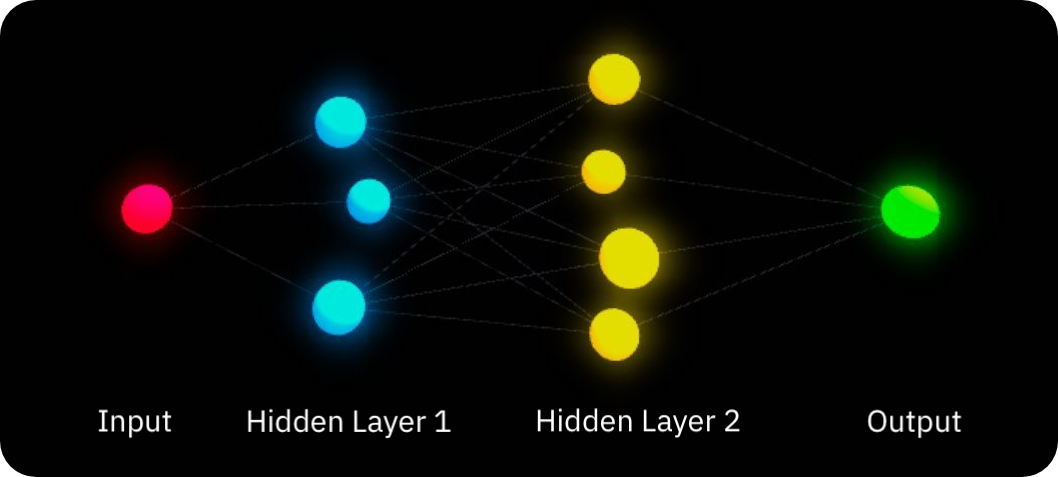}
\caption{The visualization of the neural network with three neurons in the first hidden layer (blue) and four neurons in the second hidden layer (yellow). The red node and the green node are input and output, respectively.}
\label{fig_2}
\vspace{-0.2cm}
\end{figure}

As shown in Figure~\ref{fig_2}, the NN is visualized using the node-link approach: (a) Nodes, shown as glowing spheres, represent neurons in the network. Blue nodes represent the first hidden layer; yellow nodes represent the second hidden layer. (b) Links, shown as thin white lines between nodes, represent the weights ($W_i$) between two neurons, with their transparency reflecting the magnitude of the weight: the smaller the weight, the more transparent the link. Limited by the computing power of the VR headset, also for simplicity purposes, we render the input (48 × 48) as a single node in red, the output (seven categories) as a single node in green.

We use a force-directed graph to represent the network. Specifically, the graph floats in a zero-gravity environment, with no energy loss. There are two types of forces between every two nodes ($i,j$), the attractive ($\text{FA}_{ij}$) and repulsive ($\text{FR}_{ij}$) force. The forces are defined as follows, where $k_a$ and $k_r$ being adjustable coefficients and $W_{ij}$ being the real-time weight between $i$ and $j$:
\begin{align*}
    \overrightarrow{\text{FA}}_{ij} &= k_a \times \overrightarrow{W}_{ij} & \overrightarrow{\text{FR}}_{ij} &= \frac{k_r}{\text{distance}^2(i,j)}
\end{align*}
For node pairs ($i,j$) in the same layer, there is no link and weight between them; thus, we use uniform weights with $|W_{ij}|=1$ to calculate their attraction. We normalized the value of weights ($W_{ij}$) to mitigate the impact when weights vary dramatically among neurons.

\subsection{Sonification} \label{soni}
Finding suitable mappings between the space of data and the space of sounds is conceptual~\cite{774840}. For simplicity, we map the values of validation \textit{Accuracy} and \textit{Loss} in each epoch directly to the frequency of sine wave oscillators, using a Unity plug-in \textit{Chunity}~\cite{atherton2018chunity}. Our system plays the sonification of \textit{Accuracy} on both channels by default, but the user can choose to listen to both \textit{Accuracy} and \textit{Loss} at the same time (with \textit{Accuracy} on the right channel, \textit{Loss} on the left), or only the sound of \textit{Loss} on both channels.

\begin{figure*}[!ht]
\centering
\includegraphics[width=\textwidth]{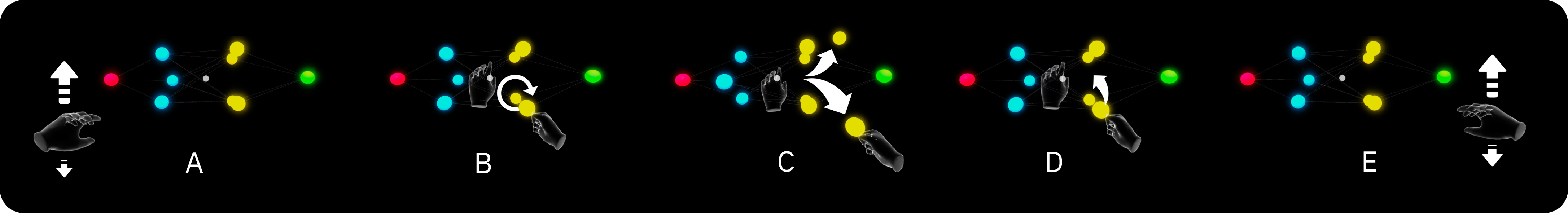}
\caption{Tuning momentum and learning rate (A, E) through hand movements; Exploring different weights by dragging nodes around (B); Updating the number of neurons by pulling new nodes out of the center (C) or dumping old nodes into the center (D).}
\label{fig_3}
\vspace{-0.2cm}
\end{figure*}

\subsection{Interaction}
% \subsection{Interacting with \AIive{}}

% Zhuoyue: comment out to shorten the length
% As the first step in exploring the concept of representing AI as a creature, we chose the most basic NN – the multi-layered perceptron~\cite{10.5555/541500}, which consists of three types of layers: input layer, hidden layer, and output layer. Once the training begins, the graph visualization will constantly move, rotate, and reshape due to the weights' changes, which affect the forces between nodes. 
\AIive{} supports three types of interactions to update the model's number of neurons, learning rate \& momentum, and weights of a single neuron (Figure~\ref{fig_3}).

\vspace{1mm}
\subsubsection{Number of Neurons} \label{num_neurons}
To update the number of neurons, the user can move the left hand close to the graph's center of mass ($<$ 5-unit distance in Unity), which would trigger the appearance of a small sphere, indicating the graph is paused. A ``Paused" message will be sent to the backend to stop the training. After that, the user can use the right hand to approach one of the hidden layers. Once the right hand enters the threshold of 3-unit distance, a small sphere in the respective layer's color would appear at its center, highlighting that the layer will be manipulated. If the user moves the right hand further closer to the layer's center of mass ($<$ 1-unit distance), a new node will be generated, and the user can drag it to the desired position. To delete a node, the user can drag it back to its layer's center. Once they have done updating, the user can put the left hand down, and the training would restart with the updated number of neurons.

% \vspace{1mm}
\subsubsection{Learning Rate and Momentum}
Once the training begins, the user can adjust the learning rate by placing the right hand in the mid-air and the left hand close to the left ear. A white sphere would then appear at the network's center, indicating the training is paused. The user can then adjust the magnitude of learning rates by lifting (increasing) or dropping (decreasing) the right hand. The new learning rate's real-time value will be sonified in the same way as \textit{Accuracy} and \textit{Loss} described in Section~\ref{soni}. Likewise, the user can adjust the momentum by placing the right hand close to the right ear and lifting/dropping the left hand. Once the user finishes updating and puts both hands down, the training would resume with updated hyperparameters.

% \vspace{1mm}
\subsubsection{Weights of a Single Neuron}
We also investigate the idea of updating weights of neurons manually, as opposed to using gradient descent. Similar to the way of updating the number of neurons in Section~\ref{num_neurons}, the user can pause the training with the left hand, then drag any existing node in the graph with the right hand. Since the operation would change the distances between the dragged node and the nodes that it connects with, the weights associated with those nodes will be updated and sent to the backend for evaluation. To guide the user on selecting the desired position, the real-time sonifications of \textit{Accuracy} and \textit{Loss} are provided, the user can move the node around and find the sweet spot where the pitch of \textit{Accuracy} sounds high (or low, in the case of \textit{Loss}) and release the node. After the user puts down the left hand, the training would resume with the updated weights.

%zhuoyue: don't detelte the following! it's the old layout
% \begin{figure}[!htb]
% \centering
% \includegraphics[width=\linewidth]{./pic/3-black-rounded-93.png}
% \caption{Tuning momentum and learning rate (A, C) through hand movements; Exploring different weights by dragging nodes around (B); Updating the number of neurons by pulling new nodes out of the center (D) or dumping old nodes into the center (E).}
% \label{fig_3}
% \end{figure}

\section{Implementation}
\subsection{Apparatus}
The NN model (back-end) runs in the python terminal of a laptop, the 3D visualization and sonification (front-end) was built as a Unity~\cite{unity} project into the Oculus Quest VR headset~\cite{oculus}. The communication between them was accomplished through TCP connections~\cite{Transmission_Control_Protocol} wirelessly.
% over the same wireless local network.

\subsection{Model}
We deployed our system on a simple fully-connected NN in python, based on an intro to machine learning course’s materials~\cite{grosse}. The dataset provided to the model is a subset of the Toronto Faces Dataset (TFD)~\cite{susskind2010toronto}, with 3374, 419, and 385 grayscale images from TFD as the training, validation, and testing set, respectively. The network has two hidden layers, each with an adjustable number of neurons. The structure of the network, as well as cross-entropy loss ($\mathcal{L}$), are shown as follows, where $\mathbf{t}, \mathbf{y}, \mathbf{h}_1,\mathbf{h}_2,\mathbf{x}$ represents the targets, outputs, the first hidden layer, the second hidden layer, and inputs, respectively:
\begin{gather*}
    \mathbf{z}_1=\mathbf{W}_1\mathbf{x}+\mathbf{b}_1 \qquad \mathbf{h}_1= \text{ReLU}(\mathbf{z}_1)\\ 
    \mathbf{z}_2=\mathbf{W}_2\mathbf{h}_1+\mathbf{b}_2  \qquad \mathbf{h}_2= \text{ReLU}(\mathbf{z}_2) \\
     \mathbf{z}_3=\mathbf{W}_3\mathbf{h}_2+\mathbf{b}_3 \qquad \mathbf{y} = \text{Softmax}(\mathbf{z}_3) \\ \mathcal{L} = -\mathbf{t}^{\intercal}(\log \mathbf{y})
\end{gather*}
In each training step ($i$), weights ($W_i$) are updated according to the Stochastic Gradient Descent (SGD)~\cite{hinton2012neural,ruder2016overview} on the cross-entropy loss ($\mathcal{L}$) with adjustable batch size, momentum ($\mu$) and learning rate ($\varepsilon$), where $\frac{\partial \mathcal{L}'}{\partial {W}'_i}$ represent the gradient from the previous step, $\frac{\partial \mathcal{L}}{\partial {W}_i}$ represents the current gradient:
\begin{equation*}
    W_i \leftarrow W_i + \mu \frac{\partial \mathcal{L}'}{\partial {W}'_i} - \varepsilon \frac{\partial \mathcal{L}}{\partial {W}_i}
\end{equation*}
After each epoch ($\text{input\_size}/\text{batch\_size}$ steps), the \textit{Accuracy} and \textit{Loss} of the model's performance on validation sets are calculated and sent back to the VR headset, together with the new weights ($W_i$) among all neurons.
% $\frac{\text{input\_size}}{\text{batch\_size}}$

\section{Discussion and Future Work}
Our preliminary implementation and exploration with \AIive{} point out several promising future directions: (1) \AIive{} may potentially help users learn and understand neural networks' training concepts. Therefore, we plan to improve the system and conduct comprehensive studies to evaluate its educational benefits; (2) Our system only supports simple neural networks due to the limited computing power of VR systems. Modern deep learning models typically consist of tens or hundreds of thousands of neurons~\cite{lecun_deep_2015}, it would be interesting to investigate suitable designs for users to interact with a larger number of neurons in VR environments; (3) Since prior work has found different sounds could evoke different emotional states of the listener~\cite{10.3389/fcomm.2020.00046}, it is interesting to explore how different sonification in terms of timbres, pitches, and complexity would affect the user experience.

% Zhuoyue: The following point has been mentioned in (1)
% Moreover, it would be interesting to investigate how data sonification could facilitate the understanding of the fundamental concepts neural network training.

\section{Conclusion}
This paper presents \AIive{}, an interactive representation that brings AI ``alive", it leverages visual and auditory feedback to provide an artistic and intuitive way to interact with AI. Since the system does not rely on numerical values or scientific graphs, it could potentially reach a broad audience. While \AIive{} is still an early implementation, we hope to share the core idea of representing AI through the combination of 3D visualizations, sonification, and direct manipulation in VR to the broader community. We look forward to stimulating interesting conversations and to eliciting useful feedback for our future development on \AIive{}. 

\section*{Acknowledgment}
The authors would like to thank the anonymous reviewers for their reviews and suggestions, Prof. Chris Chafe and Prof. Tovi Grossman for their early feedbacks, and Prof. Fernanda Viégas, Dr. Sławomir Tadeja, Fengyuan Zhu, Jiahe Lyu, Kaiqu Liang, Chenxinran Shen for their comments.

\bibliographystyle{IEEEtran}
\bibliography{AIive}

% Generated by IEEEtran.bst, version: 1.14 (2015/08/26)
\begin{thebibliography}{10}
\providecommand{\url}[1]{#1}
\csname url@samestyle\endcsname
\providecommand{\newblock}{\relax}
\providecommand{\bibinfo}[2]{#2}
\providecommand{\BIBentrySTDinterwordspacing}{\spaceskip=0pt\relax}
\providecommand{\BIBentryALTinterwordstretchfactor}{4}
\providecommand{\BIBentryALTinterwordspacing}{\spaceskip=\fontdimen2\font plus
\BIBentryALTinterwordstretchfactor\fontdimen3\font minus
  \fontdimen4\font\relax}
\providecommand{\BIBforeignlanguage}[2]{{%
\expandafter\ifx\csname l@#1\endcsname\relax
\typeout{** WARNING: IEEEtran.bst: No hyphenation pattern has been}%
\typeout{** loaded for the language `#1'. Using the pattern for}%
\typeout{** the default language instead.}%
\else
\language=\csname l@#1\endcsname
\fi
#2}}
\providecommand{\BIBdecl}{\relax}
\BIBdecl

\bibitem{zhang2021ai}
\BIBentryALTinterwordspacing
D.~Zhang, S.~Mishra, E.~Brynjolfsson, J.~Etchemendy, D.~Ganguli, B.~Grosz,
  T.~Lyons, J.~Manyika, J.~C. Niebles, M.~Sellitto, Y.~Shoham, J.~Clark, and
  R.~Perrault, ``{The AI Index 2021 Annual Report},'' 2021. [Online].
  Available: \url{https://arxiv.org/abs/2103.06312}
\BIBentrySTDinterwordspacing

\bibitem{10.1007/978-3-319-27857-5_77}
\BIBentryALTinterwordspacing
A.~W. Harley, ``{{An Interactive Node-Link Visualization of Convolutional
  Neural Networks}},'' in \emph{Advances in Visual Computing}, G.~Bebis,
  R.~Boyle, B.~Parvin, D.~Koracin, I.~Pavlidis, R.~Feris, T.~McGraw, M.~Elendt,
  R.~Kopper, E.~Ragan, Z.~Ye, and G.~Weber, Eds.\hskip 1em plus 0.5em minus
  0.4em\relax Cham: Springer International Publishing, 2015, pp. 867--877.
  [Online]. Available: \url{https://doi.org/10.1007/978-3-319-27857-5_77}
\BIBentrySTDinterwordspacing

\bibitem{9319101}
\BIBentryALTinterwordspacing
M.~Bellgardt, C.~Scheiderer, and T.~W. Kuhlen, ``{An Immersive Node-Link
  Visualization of Artificial Neural Networks for Machine Learning Experts},''
  in \emph{{2020 IEEE International Conference on Artificial Intelligence and
  Virtual Reality (AIVR)}}, 2020, pp. 33--36. [Online]. Available:
  \url{https://doi.org/10.1109/AIVR50618.2020.00015}
\BIBentrySTDinterwordspacing

\bibitem{pmlr-v123-herrmann20a}
\BIBentryALTinterwordspacing
V.~Herrmann, ``{Visualizing and sonifying how an artificial ear hears music},''
  in \emph{{Proceedings of the NeurIPS 2019 Competition and Demonstration
  Track}}, ser. Proceedings of Machine Learning Research, H.~J. Escalante and
  R.~Hadsell, Eds., vol. 123.\hskip 1em plus 0.5em minus 0.4em\relax PMLR,
  08--14 Dec 2020, pp. 192--202. [Online]. Available:
  \url{https://proceedings.mlr.press/v123/herrmann20a.html}
\BIBentrySTDinterwordspacing

\bibitem{Zell1994}
\BIBentryALTinterwordspacing
A.~Zell, N.~Mache, R.~H{\"u}bner, G.~Mamier, M.~Vogt, M.~Schmalzl, and K.-U.
  Herrmann, \emph{SNNS (Stuttgart Neural Network Simulator)}.\hskip 1em plus
  0.5em minus 0.4em\relax Boston, MA: Springer US, 1994, pp. 165--186.
  [Online]. Available: \url{https://doi.org/10.1007/978-1-4615-2736-7_9}
\BIBentrySTDinterwordspacing

\bibitem{8371286}
\BIBentryALTinterwordspacing
F.~Hohman, M.~Kahng, R.~Pienta, and D.~H. Chau, ``{Visual Analytics in Deep
  Learning: An Interrogative Survey for the Next Frontiers},'' \emph{IEEE
  Transactions on Visualization and Computer Graphics}, vol.~25, no.~8, pp.
  2674--2693, 2019. [Online]. Available:
  \url{https://doi.org/10.1109/TVCG.2018.2843369}
\BIBentrySTDinterwordspacing

\bibitem{8400040}
\BIBentryALTinterwordspacing
F.~K. Došilović, M.~Brčić, and N.~Hlupić, ``{Explainable artificial
  intelligence: A survey},'' in \emph{2018 41st International Convention on
  Information and Communication Technology, Electronics and Microelectronics
  (MIPRO)}, 2018, pp. 0210--0215. [Online]. Available:
  \url{https://doi.org/10.23919/MIPRO.2018.8400040}
\BIBentrySTDinterwordspacing

\bibitem{10.1145/3236009}
\BIBentryALTinterwordspacing
R.~Guidotti, A.~Monreale, S.~Ruggieri, F.~Turini, F.~Giannotti, and
  D.~Pedreschi, ``{A Survey of Methods for Explaining Black Box Models},''
  \emph{ACM Comput. Surv.}, vol.~51, no.~5, Aug. 2018. [Online]. Available:
  \url{https://doi.org/10.1145/3236009}
\BIBentrySTDinterwordspacing

\bibitem{4287241}
\BIBentryALTinterwordspacing
D.~A. Bowman and R.~P. McMahan, ``{Virtual Reality: How Much Immersion Is
  Enough?}'' \emph{Computer}, vol.~40, no.~7, pp. 36--43, July 2007. [Online].
  Available: \url{https://doi.org/10.1109/MC.2007.257}
\BIBentrySTDinterwordspacing

\bibitem{SYLAIOU2010243}
\BIBentryALTinterwordspacing
S.~Sylaiou, K.~Mania, A.~Karoulis, and M.~White, ``{Exploring the relationship
  between presence and enjoyment in a virtual museum},'' \emph{International
  Journal of Human-Computer Studies}, vol.~68, no.~5, pp. 243--253, 2010.
  [Online]. Available:
  \url{https://www.sciencedirect.com/science/article/pii/S1071581909001761}
\BIBentrySTDinterwordspacing

\bibitem{fazi2016feed}
\BIBentryALTinterwordspacing
M.~B. Fazi, ``{Feed-Forward: On the Future of Twenty-First-Century Media},''
  2016. [Online]. Available:
  \url{https://press.uchicago.edu/ucp/books/book/chicago/F/bo19211873.html}
\BIBentrySTDinterwordspacing

\bibitem{kramer2010sonification}
\BIBentryALTinterwordspacing
G.~Kramer, B.~Walker, T.~Bonebright, P.~Cook, J.~H. Flowers, N.~Miner, and
  J.~Neuhoff, ``{Sonification report: Status of the field and research
  agenda},'' 2010. [Online]. Available:
  \url{https://digitalcommons.unl.edu/psychfacpub/444/}
\BIBentrySTDinterwordspacing

\bibitem{Papachristodoulou2014SonificationOL}
\BIBentryALTinterwordspacing
P.~Papachristodoulou, A.~Betella, and P.~Verschure, ``{Sonification of Large
  Datasets in a 3D Immersive Environment: A Neuroscience Case Study},'' in
  \emph{{ACHI 2014}}, 2014. [Online]. Available:
  \url{http://www.thinkmind.org/download.php?articleid=achi_2014_2_20_20146}
\BIBentrySTDinterwordspacing

\bibitem{+2020}
\BIBentryALTinterwordspacing
A.~Sudmann, Ed., \emph{{The Democratization of Artificial Intelligence: Net
  Politics in the Era of Learning Algorithms}}.\hskip 1em plus 0.5em minus
  0.4em\relax transcript Verlag, 2020. [Online]. Available:
  \url{https://doi.org/10.14361/9783839447192}
\BIBentrySTDinterwordspacing

\bibitem{10.1145/3379336.3381474}
\BIBentryALTinterwordspacing
D.~Wang, P.~Ram, D.~K.~I. Weidele, S.~Liu, M.~Muller, J.~D. Weisz, A.~Valente,
  A.~Chaudhary, D.~Torres, H.~Samulowitz, and L.~Amini, ``{AutoAI: Automating
  the End-to-End AI Lifecycle with Humans-in-the-Loop},'' in \emph{{Proceedings
  of the 25th International Conference on Intelligent User Interfaces
  Companion}}, ser. IUI '20.\hskip 1em plus 0.5em minus 0.4em\relax New York,
  NY, USA: Association for Computing Machinery, 2020, p. 77–78. [Online].
  Available: \url{https://doi.org/10.1145/3379336.3381474}
\BIBentrySTDinterwordspacing

\bibitem{candel2016deep}
\BIBentryALTinterwordspacing
A.~Candel, V.~Parmar, E.~LeDell, and A.~Arora, ``{Deep learning with H2O},''
  \emph{H2O. ai Inc}, pp. 1--21, 2016. [Online]. Available:
  \url{https://web.archive.org/web/20211018215342/https://www.h2o.ai/}
\BIBentrySTDinterwordspacing

\bibitem{smilkov2017directmanipulation}
\BIBentryALTinterwordspacing
D.~Smilkov, S.~Carter, D.~Sculley, F.~B. Viégas, and M.~Wattenberg,
  ``{Direct-Manipulation Visualization of Deep Networks},'' 2017. [Online].
  Available: \url{https://arxiv.org/abs/1708.03788}
\BIBentrySTDinterwordspacing

\bibitem{8440049}
\BIBentryALTinterwordspacing
M.~Kahng, N.~Thorat, D.~H. Chau, F.~B. Viégas, and M.~Wattenberg, ``{GAN Lab:
  Understanding Complex Deep Generative Models using Interactive Visual
  Experimentation},'' \emph{IEEE Transactions on Visualization and Computer
  Graphics}, vol.~25, no.~1, pp. 310--320, Jan 2019. [Online]. Available:
  \url{https://doi.org/10.1109/TVCG.2018.2864500}
\BIBentrySTDinterwordspacing

\bibitem{10.1145/3379337.3415878}
\BIBentryALTinterwordspacing
D.~Hayatpur, H.~Xia, and D.~Wigdor, ``{DataHop: Spatial Data Exploration in
  Virtual Reality},'' in \emph{{Proceedings of the 33rd Annual ACM Symposium on
  User Interface Software and Technology}}, ser. UIST '20.\hskip 1em plus 0.5em
  minus 0.4em\relax New York, NY, USA: Association for Computing Machinery,
  2020, p. 818–828. [Online]. Available:
  \url{https://doi.org/10.1145/3379337.3415878}
\BIBentrySTDinterwordspacing

\bibitem{tadeja_seshadri_kristensson_2020}
\BIBentryALTinterwordspacing
S.~Tadeja, P.~Seshadri, and P.~Kristensson, ``{AeroVR: An immersive
  visualisation system for aerospace design and digital twinning in virtual
  reality},'' \emph{The Aeronautical Journal}, vol. 124, no. 1280, p.
  1615–1635, 2020. [Online]. Available:
  \url{https://doi.org/10.1017/aer.2020.49}
\BIBentrySTDinterwordspacing

\bibitem{vanhorn2019deep}
\BIBentryALTinterwordspacing
K.~C. VanHorn, M.~Zinn, and M.~C. Cobanoglu, ``{Deep Learning Development
  Environment in Virtual Reality},'' 2019. [Online]. Available:
  \url{https://arxiv.org/abs/1906.05925}
\BIBentrySTDinterwordspacing

\bibitem{6871829}
\BIBentryALTinterwordspacing
M.~Chandramouli, M.~Zahraee, and C.~Winer, ``{A fun-learning approach to
  programming: An adaptive Virtual Reality (VR) platform to teach programming
  to engineering students},'' in \emph{{IEEE International Conference on
  Electro/Information Technology}}, June 2014, pp. 581--586. [Online].
  Available: \url{https://doi.org/10.1109/EIT.2014.6871829}
\BIBentrySTDinterwordspacing

\bibitem{9419195}
\BIBentryALTinterwordspacing
J.~Pirker, J.~Kopf, A.~Kainz, A.~Dengel, and B.~Buchbauer, ``{The Potential of
  Virtual Reality for Computer Science Education -Engaging Students through
  Immersive Visualizations},'' in \emph{{2021 IEEE Conference on Virtual
  Reality and 3D User Interfaces Abstracts and Workshops (VRW)}}, March 2021,
  pp. 297--302. [Online]. Available:
  \url{https://doi.org/10.1109/VRW52623.2021.00060}
\BIBentrySTDinterwordspacing

\bibitem{9320876}
\BIBentryALTinterwordspacing
S.~Paul and S.~Hamad, ``{The Role of Virtual Reality in Story telling and Data
  Visualization for motivating students in learning programming},'' in
  \emph{{2020 Seventh International Conference on Information Technology Trends
  (ITT)}}, Nov 2020, pp. 169--173. [Online]. Available:
  \url{https://doi.org/10.1109/ITT51279.2020.9320876}
\BIBentrySTDinterwordspacing

\bibitem{10.1145/1978942.1979193}
\BIBentryALTinterwordspacing
N.~Diakopoulos, F.~Kivran-Swaine, and M.~Naaman, ``{Playable Data:
  Characterizing the Design Space of Game-y Infographics},'' in
  \emph{{Proceedings of the SIGCHI Conference on Human Factors in Computing
  Systems}}, ser. CHI '11.\hskip 1em plus 0.5em minus 0.4em\relax New York, NY,
  USA: Association for Computing Machinery, 2011, p. 1717–1726. [Online].
  Available: \url{https://doi.org/10.1145/1978942.1979193}
\BIBentrySTDinterwordspacing

\bibitem{10.1145/3170427.3188537}
\BIBentryALTinterwordspacing
P.~Millais, S.~L. Jones, and R.~Kelly, ``{Exploring Data in Virtual Reality:
  Comparisons with 2D Data Visualizations},'' in \emph{{Extended Abstracts of
  the 2018 CHI Conference on Human Factors in Computing Systems}}, ser. CHI EA
  '18.\hskip 1em plus 0.5em minus 0.4em\relax New York, NY, USA: Association
  for Computing Machinery, 2018, p. 1–6. [Online]. Available:
  \url{https://doi.org/10.1145/3170427.3188537}
\BIBentrySTDinterwordspacing

\bibitem{de1999sonification}
\BIBentryALTinterwordspacing
A.~de~Campo, \emph{{Sonification of social data}}.\hskip 1em plus 0.5em minus
  0.4em\relax Ann Arbor, MI: Michigan Publishing, University of Michigan
  Library, 1999. [Online]. Available:
  \url{http://hdl.handle.net/2027/spo.bbp2372.1999.397}
\BIBentrySTDinterwordspacing

\bibitem{10.1145/3429290.3429307}
\BIBentryALTinterwordspacing
S.~Nath, ``{Hear Her Fear: Data Sonification for Sensitizing Society on Crime
  Against Women in India},'' in \emph{{IndiaHCI '20: Proceedings of the 11th
  Indian Conference on Human-Computer Interaction}}, ser. IndiaHCI 2020.\hskip
  1em plus 0.5em minus 0.4em\relax New York, NY, USA: Association for Computing
  Machinery, 2020, p. 86–91. [Online]. Available:
  \url{https://doi.org/10.1145/3429290.3429307}
\BIBentrySTDinterwordspacing

\bibitem{10.1145/1873951.1874219}
\BIBentryALTinterwordspacing
R.~Valenti, A.~Jaimes, and N.~Sebe, ``{Sonify Your Face: Facial Expressions for
  Sound Generation},'' in \emph{{Proceedings of the 18th ACM International
  Conference on Multimedia}}, ser. MM '10.\hskip 1em plus 0.5em minus
  0.4em\relax New York, NY, USA: Association for Computing Machinery, 2010, p.
  1363–1372. [Online]. Available:
  \url{https://doi.org/10.1145/1873951.1874219}
\BIBentrySTDinterwordspacing

\bibitem{r2019hearing}
\BIBentryALTinterwordspacing
W.~R~Michael, A.~Kalra, and B.~N. Walker, ``{Hearing artificial intelligence:
  Sonification guidelines \& results from a case-study in melanoma
  diagnosis}.''\hskip 1em plus 0.5em minus 0.4em\relax Georgia Institute of
  Technology, 2019. [Online]. Available:
  \url{https://doi.org/10.21785/icad2019.021}
\BIBentrySTDinterwordspacing

\bibitem{6126903}
\BIBentryALTinterwordspacing
G.~Dublon, L.~S. Pardue, B.~Mayton, N.~Swartz, N.~Joliat, P.~Hurst, and J.~A.
  Paradiso, ``{DoppelLab: Tools for exploring and harnessing multimodal sensor
  network data},'' in \emph{{SENSORS, 2011 IEEE}}, Oct 2011, pp. 1612--1615.
  [Online]. Available: \url{https://doi.org/10.1109/ICSENS.2011.6126903}
\BIBentrySTDinterwordspacing

\bibitem{774840}
\BIBentryALTinterwordspacing
H.~Kaper, E.~Wiebel, and S.~Tipei, ``{Data sonification and sound
  visualization},'' \emph{Computing in Science and Engineering}, vol.~1, no.~4,
  pp. 48--58, July 1999. [Online]. Available:
  \url{https://doi.org/10.1109/5992.774840}
\BIBentrySTDinterwordspacing

\bibitem{atherton2018chunity}
\BIBentryALTinterwordspacing
J.~Atherton and G.~Wang, ``{Chunity: Integrated Audiovisual Programming in
  Unity},'' in \emph{{NIME}}, 2018, pp. 102--107. [Online]. Available:
  \url{https://www.nime.org/proceedings/2018/nime2018_paper0024.pdf}
\BIBentrySTDinterwordspacing

\bibitem{unity}
\BIBentryALTinterwordspacing
Unity, ``{{Unity Real-Time Development Platform | 3D, 2D VR and AR Engine}}.''
  [Online]. Available:
  \url{https://web.archive.org/web/20211020091658/https://unity.com/}
\BIBentrySTDinterwordspacing

\bibitem{oculus}
\BIBentryALTinterwordspacing
Oculus, ``{{Oculus Quest VR headset}}.'' [Online]. Available:
  \url{{https://web.archive.org/web/20211018111646/https://www.oculus.com/quest/features/}}
\BIBentrySTDinterwordspacing

\bibitem{Transmission_Control_Protocol}
Wikipedia, ``{{Transmission Control Protocol}} --- {{W}}ikipedia{{,}} the free
  encyclopedia,''
  \url{https://en.wikipedia.org/wiki/Transmission_Control_Protocol}, 2021,
  [Online; accessed 20-October-2021].

\bibitem{grosse}
\BIBentryALTinterwordspacing
R.~Grosse, ``{CSC 311 Fall 2020: Introduction to Machine Learning}.'' [Online].
  Available:
  \url{https://web.archive.org/web/20211020015616/https://www.cs.toronto.edu/~rgrosse/courses/csc311_f20/}
\BIBentrySTDinterwordspacing

\bibitem{susskind2010toronto}
\BIBentryALTinterwordspacing
J.~M. Susskind, A.~K. Anderson, and G.~E. Hinton, ``{The toronto face
  database},'' \emph{Department of Computer Science, University of Toronto,
  Toronto, ON, Canada, Tech. Rep}, vol.~3, 2010. [Online]. Available:
  \url{https://zhuoyuelyu.github.io/AIive/toronto_face.npz}
\BIBentrySTDinterwordspacing

\bibitem{hinton2012neural}
\BIBentryALTinterwordspacing
G.~Hinton, N.~Srivastava, and K.~Swersky, ``{Neural networks for machine
  learning lecture 6a overview of mini-batch gradient descent},'' \emph{Cited
  on}, vol.~14, no.~8, p.~2, 2012. [Online]. Available:
  \url{https://web.archive.org/web/20211018165553/https://www.cs.toronto.edu/~tijmen/csc321/slides/lecture_slides_lec6.pdf}
\BIBentrySTDinterwordspacing

\bibitem{ruder2016overview}
\BIBentryALTinterwordspacing
S.~Ruder, ``{An overview of gradient descent optimization algorithms},''
  \emph{arXiv preprint arXiv:1609.04747}, 2016. [Online]. Available:
  \url{https://arxiv.org/pdf/1609.04747.pdf}
\BIBentrySTDinterwordspacing

\bibitem{lecun_deep_2015}
\BIBentryALTinterwordspacing
Y.~{LeCun}, Y.~Bengio, and G.~Hinton, ``{Deep learning},'' vol. 521, no. 7553,
  pp. 436--444. [Online]. Available: \url{https://doi.org/10.1038/nature14539}
\BIBentrySTDinterwordspacing

\bibitem{10.3389/fcomm.2020.00046}
\BIBentryALTinterwordspacing
N.~Sawe, C.~Chafe, and J.~Treviño, ``Using data sonification to overcome
  science literacy, numeracy, and visualization barriers in science
  communication,'' \emph{Frontiers in Communication}, vol.~5, p.~46, 2020.
  [Online]. Available:
  \url{https://www.frontiersin.org/article/10.3389/fcomm.2020.00046}
\BIBentrySTDinterwordspacing

\end{thebibliography}
\end{document}